\documentclass[twocolumn]{aastex631}

\usepackage{booktabs} % Required for \addlinespace
\usepackage{xcolor}
\usepackage{amsmath}

\begin{document}

\title{Hierarchical searches for subsolar-mass binaries and the third-generation gravitational wave detector era}

\correspondingauthor{Kanchan Soni}
\email{ksoni01@syr.edu}

\author[0000-0001-8051-7883]{Kanchan Soni}
\affiliation{Department of Physics, Syracuse University, Syracuse, New York 13244, USA}
\affiliation{Inter-University Centre for Astronomy and Astrophysics, Pune 411007, India}

\author[0000-0002-1850-4587]{Alexander H. Nitz}
\affiliation{Department of Physics, Syracuse University, Syracuse, New York 13244, USA}




\begin{abstract}

% shorter version (less than 250 words for the submission)
The detection of gravitational waves (GWs) from coalescing compact binaries has become routine with ground-based detectors like LIGO and Virgo. However, beyond standard sources such as binary black holes and neutron stars and neutron star black holes, no exotic sources revealing new physics have been discovered. Detecting ultra-compact objects, such as subsolar mass (SSM) compact objects, offers a promising opportunity to explore diverse astrophysical populations. However, searching for these objects using standard matched-filtering techniques is computationally intensive due to the dense parameter space involved. This increasing computational demand not only challenges current search methodologies but also poses significant obstacles for third-generation (3G) ground-based GW detectors. In the 3G era, signals may last tens of minutes, and detection rates could reach one per minute, requiring efficient search strategies to manage the computational load of long-duration signals. In this paper, we demonstrate a hierarchical search strategy designed to address the challenges of searching for long-duration signals, such as those from SSM compact binaries, and the anticipated issues with 3G detectors. We show that by adopting optimization techniques in a two-stage hierarchical approach, we can efficiently search for the SSM compact object in the current LIGO detectors. Our preliminary results show that conducting matched filtering at a lower frequency of 35 Hz improves the signal-to-noise ratio by 6\% and enhances the detection volume by 10-20\%, compared to the standard two-detector PyCBC search. This improvement is achieved while reducing computational costs by a factor of 2.5.

\end{abstract}


\section{Introduction} \label{sec:intro}

The field of gravitational-wave astronomy has been rapidly expanding ever since the detection of the first binary black hole merger GW150914~\citep{PhysRevLett.116.061102}. To date, nearly 90 gravitational wave (GW) sources are cataloged by LIGO-Virgo-KAGRA (LVK) Collaboration, including dozens of binary black holes, two binary neutron stars, and three neutron star-black hole mergers~\citep{gwtc3}. Additional GW sources were independently cataloged~\citep{nitz_ogc3,venumadhav_2022_o3a,ogc4_Nitz_2023,mehta2024newbinaryblackhole,wadekar2023newblackholemergers} using publicly available data~\citep{KAGRA:2023pio,LIGOScientific:2019lzm}. The third observation (O3) run of Advanced LIGO~\citep{asi} and Advanced Virgo~\citep{advancedvirgo} lead to the detection of compact objects within sub-3 $\rm M_{\odot}$ range with GW190814~\citep{gw190814} where the secondary compact object had a mass of $\sim2.59\, \rm M_{\odot}$ and a low spin ($\le 0.07$). Several other events like GW190425, GW191219, GW200105, GW200115, and GW200210 identified during this run also had one of the component masses less than $3\, \rm M_{\odot}$. The ongoing fourth observation run detected GW230529~\citep{gw230529}. This event's primary object had a mass ranging between 2.5 and $4.5\, \rm M_{\odot}$, making it an additional compact object, likely a black hole (BH), existing within the `lower mass gap'~\citep{Bailyn_1998,Ozel_2010,Farr_2011}. Although several studies have provided insights into the mass and spin distributions of compact sources detected through current GW detectors~\citep{Abbott_2021_pop,2019MNRAS.484.4216R}, the possibility of discovering ultra-compact objects with masses less than a solar mass range remains an open question~\citep{subsolar_lvk_o3a,subsolar_lvk19,subsolar_lvk21,nitz_subsolar_2022,nitz_wang_subsolar,lvk_subsolar_mass_search_2022,Miller:2024rca}.

Subsolar mass (SSM) compact objects do not follow the standard stellar evolution pathway. These objects, if BHs, are expected to form through non-stellar evolution models and could be primordial black holes (PBHs)~\citep{PhysRevD.81.104019}. If they are neutron stars~\citep{2022NatAs...6.1444D}, they might result from non-standard supernova explosion models~\citep{müller2024minimumneutronstarmass}. Although the search for SSM black holes began quite early~\citep{Nakamura_1997,Alcock_2000}, no candidates have been found yet. Since then, numerous models have proposed various formation pathways for these sources. The most common formation mechanism of PBHs suggests their origin from the direct collapse of early, small-scale fluctuations~\citep{zeldovich_1967,hawking_1971} due to certain features of the inflationary potential. Additionally, there are alternative formation channels where PBHs emerge from phase transitions~\citep{Byrnes_2018} in the early universe or through the collapse of topological defects like cosmic strings~\citep{HAWKING1989237,PhysRevD.43.1106,Cheng_Hong-bo_1996}.

Studies show that a small fraction of dark matter could be due to PBHs~\citep{PhysRevD.81.104019,Carr_2021}. While many cosmological investigations have ruled out their existence at extremely low masses~\citep{Sasaki_2018}, exploration continues in a mass range spanning several orders of magnitude. If these black holes appear in a binary system, the emitted GWs can be detected through ground-based interferometers. Several studies have investigated the search for SSM black holes~\citep{subsolar_lvk19,nitz_wang_subsolar,Nitz_pbh_2021,subsolar_lvk_o3a,subsolar_lvk21}, but no significant detections have been made to date. A confirmed detection within the LIGO-Virgo frequency band would provide critical insights into the formation mechanisms of PBHs and contribute to constraining the fraction of dark matter in the universe.

The offline search for GWs from the inspiralling and merging binaries uses the matched filtering technique~\citep{Phys.Rev.D1991,svd-sathya,svd-schutz,sathya-owen,findchirp,usman_2016,hlv_gareth_pycbc}. In this method, a bank of modeled signals, or templates, is correlated with well-calibrated interferometer data~\citep{Xavier_Siemens_2004,ABADIE2010223}. However, this approach becomes computationally demanding, particularly for low-mass binaries, as the cost increases with the number of templates and the length of the signal model used as a matched filter template. To mitigate the computational challenges, suboptimal choices are often made by limiting the search parameters. For instance, searches may only filter data above 45 Hz~\citep{advligo2}, or limit the duration of the templates to nearly 512 seconds~\citep{nitz_wang_subsolar,nitz_subsolar_2022}. While these restrictions help reduce computational costs, they also reduce the sensitive volume by approximately 24\%, within which PBHs might be detected.

Observing long-duration GW signals poses significant challenges with existing search methods. The main difficulties arise from the necessity of using a very dense template bank ($\mathcal{O}(10^7)$), which significantly increases the computational cost of the matched filtering search. Furthermore, the search sensitivity can be compromised by non-stationary data, which may contain long-duration correlations that hinder current signal-vetoing techniques and statistical analyses. This non-stationarity can also impact the statistics and signal-vetoing methods used in current search pipelines. These issues are expected to become considerably more severe in the era of third-generation (3G) ground-based detectors. 3G detectors such as the Cosmic Explorer~\citep{cosmic_explorer_Abbott_2017,cosmic_explorer_white_paper} and the Einstein Telescope~\citep{Hild_2010,Punturo_2010,ET2017,Maggiore_2020,Pace_ET_2022}, are anticipated to detect binary mergers at rates two to three orders of magnitude higher than current detectors~\citep{evans2021horizonstudycosmicexplorer}. These detectors, designed to operate from very low frequencies (starting from 2 Hz), will observe signals for several minutes or hours. Due to their high sensitivity in the lower frequency band, the likelihood of detecting eccentric or precessing binaries will be higher, which will indirectly expand the template bank’s parameter space—both in dimensionality and parameter ranges—thereby increasing the computational cost of the search. Furthermore, since signals would remain in the sensitivity band for longer periods, the Earth’s rotation will reduce search sensitivity by altering the detector’s response functions and affecting matched filter statistics. Therefore, developing an efficient, cost-effective matched filtering strategy for long-duration GW signals is essential to advance the current state-of-the-art search techniques. This approach will not only mitigate computational burdens but also prepare us to effectively search for GW signals from compact binary coalescences (CBCs) with 3G detectors.

% Additionally, projected abundance of SSM black holes in the 3G era~\citep{http://dx.doi.org/10.3847/2041-8213/ac6bea,https://dx.doi.org/10.1088/1475-7516/2020/08/039} will present significant challenges in their detection. 

One approach to efficiently search for long-duration signals, such as those from SSM binaries, is implementing a hierarchical search strategy~\citep{hierarchical_2022,PhysRevD.105.103001,hierarchical_bg_2024}. In this method, a two-stage matched filtering search is performed using multiple template banks of varying densities. In the first stage, the search is conducted over coarsely sampled data using a coarse bank to identify coincident triggers that could represent true GW events. These triggers are followed up in the second stage with a finer search, focusing on the neighborhood of the parameter space identified in the first stage. This two-stage approach effectively reduces the number of matched filtering operations required for the search, significantly reducing computational time.

In this paper, we present a comprehensive analysis of real data to demonstrate how hierarchical search strategies can be effectively applied to SSM binary searches in Advanced LIGO data. We also discuss the necessary modifications to extend these techniques to generic CBC searches in the upcoming 3G detectors. We implement a two-stage hierarchical search method, as detailed in~\cite{hierarchical_2022,hierarchical_bg_2024}, specifically targeting SSM compact objects. This method allows us to optimize the search by employing different sampling rates and varying template bank densities across the two stages.

Our search focuses on SSM binaries with primary masses $m_1 \in [0.2, 10]\, M_\odot$ and secondary masses $m_2 \in [0.2, 1] \, M_\odot$, defined in the detector frame. We also consider component spins of up to 0.9 for each compact object. The parameter space for our template bank is similar to that used by the LVK collaboration~\citep{subsolar_lvk_o3a}, but we operate with different frequency settings. Specifically, our search is tuned to detect SSM candidates beginning at 35 Hz within the Advanced LIGO frequency band. By lowering the operational frequency, we aim to reduce the loss in astrophysical volume by approximately 10\%. Although this adjustment increases the density of the coarse template bank by a factor of 1.5 compared to the bank used in the PyCBC search in~\cite{subsolar_lvk_o3a}, we achieve a reduction in the computational cost of matched filtering by utilizing two stages with varying data sampling rates in our search pipeline.

\section{\label{sec:method}Method}

The matched filtering search for long-duration signals as in the case of binaries containing SSM compact objects is expensive as it requires a very dense template bank. To optimize the search, often the length of the template is reduced to a manageable duration ($\sim512$ seconds) so that search analysis can be performed. This could be enabled by performing matched filtering from 45 Hz rather than 15 Hz~\citep{gwtc3}. However, such adjustments affect the horizon distance of the binary and the expected signal-to-noise (SNR) ratio. 

The horizon distance~\citep{1987thyg.book..330T,findchirp} for an inspiraling binary is given by

\begin{equation}
    D_{\rm max} \propto \frac{\mathcal{M}^{5/6}}{\rho} \sqrt{\int^{f_{max}}_{f_{min}}\frac{f^{-7/3}}{S_{n}(f)} \, df}\,,
\end{equation}

where $f_{min}$ and $f_{max}$ are the minimum and maximum frequencies of the LIGO's sensitivity range. $S_n (f)$ is the power spectral density (PSD) of the noise in the detector and $\rho$ is expected matched filter SNR for an inspiraling binary with chirp mass $\mathcal{M}$ observed in the detector's frame. 

For a particular source of chirp mass of a few that have a fixed SNR in the detector's frame, changing the operating frequency band can affect the detectability of a signal. This means that the fractional SNR loss, as also shown in~\cite{Magee_2018}, due to a change in the operating frequency band would be

\begin{equation}\label{eq:fractional_loss_in_snr}
    F_{\rm SNR-loss} = 1 - \frac{D_{\rm max}(f_{min}\,, f_{max})}{D_{\rm max}(15 \rm Hz\,, 2048 \rm Hz)}\,,
\end{equation}

with respect to the standard operational band of Advanced LIGO for matched filtering for generic CBC search~\citep{gwtc3}. 

\begin{figure}[ht!]
    \centering
    \includegraphics[scale=.45]{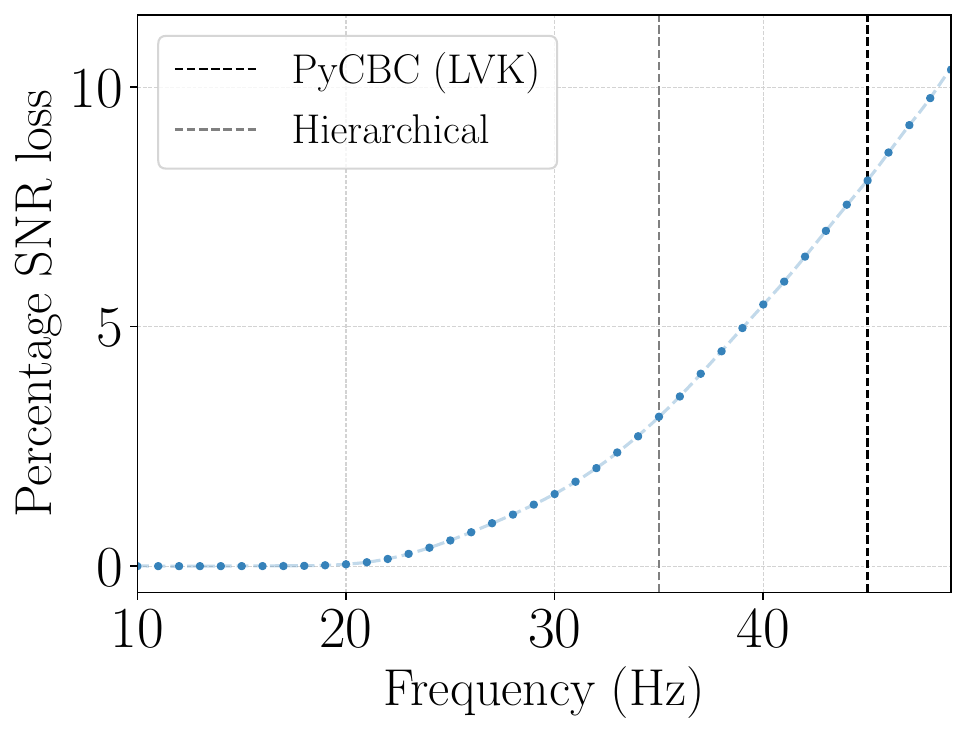}
    \caption{The percentage loss in SNR as a function of lower frequency limit ($f_{\rm min}$) used for matched filtering data, as described in Eq.~\eqref{eq:fractional_loss_in_snr}. The SNR loss increases with larger $f_{min}$ values in matched filtering. For comparison, the black dotted line represents the lower frequency limit of 45 Hz~\citep{subsolar_lvk_o3a}, while the gray line indicates the 35 Hz limit used in this work.}
    \label{fig:loss_in_snr}
\end{figure}

If we perform a search from 35 Hz, assuming the other source parameters do not change, the percentage loss in SNR is about 3.1\% for a frequency band of $35-2048$ Hz. This loss is relatively small compared to the SSM searches conducted in the $45-2048$ Hz band~\citep{subsolar_lvk_o3a}, which experience an SNR loss of about $8-9$\%. This comparison can also be seen in Fig.~\ref{fig:loss_in_snr}. Ideally, the lower frequency limit can be further lowered to match those used for a generic CBC search. However, this step will increase the computation demand and require very long-duration templates in the bank. Therefore, in our work, we select a lower frequency cutoff of 35 Hz. By making this choice, we expect the loss in astrophysical volumetric sensitivity to the inspiral stage by $\sim 10\%$ of binary coalescence, which is lower than 24\% as observed in~\cite{subsolar_lvk_o3a}.

Generating templates at lower frequencies enhances the sensitivity of the search, but it can also lead to a higher density of the template bank. This increase in density may raise the computational cost of matched filtering in traditional offline PyCBC or \textit{flat} search~\citep{usman_2016,hlv_gareth_pycbc}. However, this added computational burden can be mitigated by adopting hierarchical search strategies.

\subsection{Review of hierarchical search}

The two-detector flat search performs matched filtering over a discretely sampled data segment using a dense bank of templates and generates \textit{triggers} with SNRs ($\rho$)~\citep{usman_2016}. In contrast, a hierarchical search strategy offers a more efficient approach by enabling a multi-stage matched filtering process, where the number of templates is progressively reduced at each stage. As described in~\cite{hierarchical_2022}, this search divides a flat search into two stages: \textit{coarse} and \textit{fine}. During the coarse stage, the data is matched filtered using a \textit{coarse} template bank, which consists of sparsely placed templates. The sparseness of these templates is determined by the minimal match~\citep{owen_templatebank} at which the bank is constructed, typically set lower (below 0.97) than the value used for constructing a flat bank. 

Performing a coarse search reduces the number of matched filtering operations. This reduction is further enhanced when the data is sampled at a lower frequency (512 Hz). However, this approach comes with the trade-off of potentially lower matched filter SNR values for the resulting triggers. To address this, the SNR thresholds are lowered ($\rho = 3.5$) compared to those used in a flat search ($\rho = 4$). Given these triggers could be generated by non-Gaussian features or \textit{glitches}~\citep{ligo_burst_search_2021} present in the data, the SNRs are further down-weighted using chi-square vetoes~\citep{bruceallen,nitzsgveto}. Only those triggers that pass these vetoes are then subjected to a coincidence test~\citep{usman_2016}, during which a ranking statistic is computed to assess their significance~\citep{nitzphasetd}.

The coincident triggers obtained from the coarse search, with ranking statistics above a certain threshold (approximately 7, as used in~\cite{hierarchical_2022}), are followed up in the second stage for a finer search. In this stage, a focused search is conducted within the vicinity or neighborhood (\textit{nhbd}) of the coarse template’s parameter space. This nbhd is a region around a coarse template where the minimal match between templates within the nbhd ($\sim$10–100) and the coarse template ranges from 0.75 to 0.99. 

To avoid the computational burden of calculating the nbhd for each coarse template on the fly, a pre-computed nbhd bank is used. This bank contains all the nbhd regions and the corresponding templates for each coarse template. During the second stage search, a union of all the nbhds corresponding to the coarse triggers in each data segment is used for matched filtering. To further improve the SNR, the data sampling rate is increased to 2408 Hz, which is higher than that used in the coarse search. Triggers with SNRs above 4 that pass all chi-square tests are then subjected to a final coincidence search, compiling a list of foreground candidates.

To assess the significance of potential GW events, the false alarm rate (FAR) is estimated based on the background~\citep{usman_2016}. Unlike the flat search, which estimates the background by applying millions of time shifts to triggers from a single detector, the hierarchical search employs a hybrid approach in its second stage~\citep{hierarchical_bg_2024}. At first, a few time shifts are applied to generate coincident background triggers. Then, an exponential fit is applied to the cumulative distribution of these background triggers, and the fitted curve is used to calculate the FAR for the foreground triggers obtained in the second stage. This method of estimating the background is particularly effective for long-duration signals, as the expected background distribution tends to follow the tail of a Poisson distribution~\citep{usman_2016}.

\subsection{Template bank}\label{sec:subsolar_mass_template_bank}

We construct two aligned-spin banks-- a coarse bank at a minimal match of 0.92 and a fine bank of 0.97, using a geometric placement algorithm~\citep{geometric_placement_bank_gen} for the hierarchical search. Both banks are designed to cover parameters where $m_1$ ranges between $0.2-10 \rm~M_{\odot}$ and $m_2$ between $0.2-1.0 \rm~M_{\odot}$ in the detector frame. The dimensionless spins span up to $0.9$ for both compact objects. These bank parameter ranges are consistent with the bank used for the LVK search~\citep{subsolar_lvk_o3a,lvk_subsolar_mass_search_2022}. From here, we refer to this bank as \textit{flat bank}.

In contrast to the flat bank, where templates commence at a frequency of 45 Hz, the templates in our banks start at a frequency of 35 Hz. This choice reduces the fractional loss in the matched filter SNR, as shown in Sec.~\ref{sec:method}. As a result, even though our coarse bank is constructed at a lower minimal match, it is approximately 1.6 times the size of the flat bank. These distinctions are summarized in Table~\ref{table:table_subsolar_banks}.

\begin{table}[ht!]
\centering
\caption{Summary of the coarse and fine template banks constructed for the hierarchical search, with a comparison to the flat bank used in the LVK SSM search~\citep{subsolar_lvk_o3a}. The banks are characterized by different minimal match values, which denote the minimum match between neighboring templates, and different starting frequencies $f_0$. A lower $f_0$ results in increased template density, as demonstrated by the fine bank, even though the fine and flat banks have similar minimal match values. The coarse bank is approximately 1.6 times denser than the flat bank, primarily due to its lower starting frequency.}
\begin{tabular}{
l@{\hspace{20pt}}
c@{\hspace{20pt}}
c@{\hspace{20pt}}
c@{\hspace{5pt}}}
\hline
% \hline
% \addlinespace
Bank & $f_{0}$ (Hz)  & Templates & Minimal match  \\
% \addlinespace
\hline
\addlinespace
Coarse & 35 &  2,961,067 & 0.92 \\
Fine & 35 & 8,886,979 & 0.97  \\
Flat & 45 & 1,864,323 & 0.97  \\
% Coarse & 35 &  2,961,601 & 0.90 \\
\addlinespace
% \hline
\hline
\end{tabular}
\label{table:table_subsolar_banks}
\end{table}

\begin{figure}[ht!]
    \centering
    \includegraphics[scale=.35]{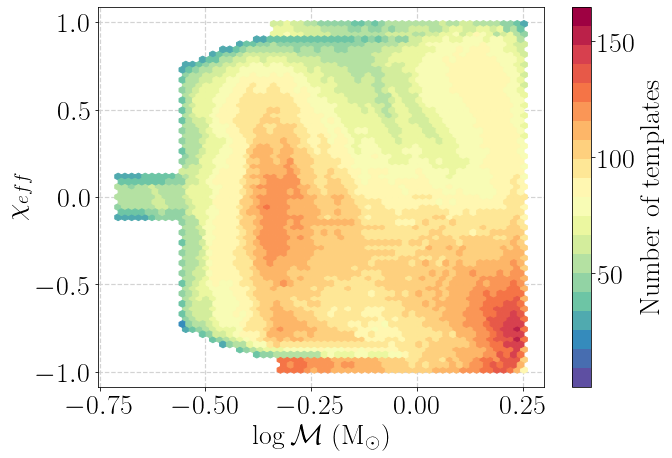}
    \caption{Figure depicting the distribution of coarse templates in the logarithm of chirp mass ($\mathcal{M}$) - effective spin ($\chi_{\text{eff}}$) plane. The color bar represents the total number of templates in the nbhd of each coarse template parameter.}
    \label{fig:subsolar_nbhd_bank}
\end{figure}

The hierarchical search requires the construction of nbhd bank for the second stage search. Therefore, we use the generated fine bank to construct the nbhd bank using the method described in Sec. IV of~\cite{hierarchical_2022}. Figure~\ref{fig:subsolar_nbhd_bank} shows the parameter space covered by the coarse templates in the chirp mass and effective spin plane. This plot shows that the distribution of templates within a nbhd is not uniform across the parameter space, primarily due to boundary effects. These effects occur because the match between neighboring templates gradually decreases as the mismatch in the $\tau_3$ mass parameter increases relative to $\tau_0$, thereby significantly extending the nbhd region along this coordinate.

% http://localhost:8883/notebooks/o3_subsolar_bank/final_banks/bank_plots.ipynb

As shown in Fig.~\ref{fig:subsolar_nbhd_bank}, the number of templates in a nbhd typically ranges from a few tens to hundreds. This can significantly affect the final background in the second stage. To improve background estimation, more noise coincidences from the coarse search need to be followed up. However, since the number of templates in the nbhds is relatively small, the search cost is not expected to be significantly higher than that of the flat search.

\begin{table*}[ht!]
\centering
\caption{Results from a two-detector analysis over data duration from April 1 to April 8, 2019, using flat and hierarchical search pipelines. Listed candidates are arranged in descending order of false alarm rate (FAR). The table also compares chirp mass ($\mathcal{M}$) and network SNR $(\hat{\rho}_{T}=\sqrt{\rho^2_{H}+\rho^2_{L}})$ for each identified candidates. The FARs of the detected events in the flat search were determined using the time-shift method, whereas those for the hierarchical search were determined by the method described in~\cite{hierarchical_bg_2024}.}
\begin{tabular}{
l@{\hspace{8pt}}
c@{\hspace{8pt}}c@{\hspace{8pt}}c@{\hspace{8pt}}c@{\hspace{12pt}}
c@{\hspace{8pt}} c@{\hspace{8pt}} c@{\hspace{8pt}}}
\hline
% \addlinespace
 & \multicolumn{3}{c}{Hierarchical}  & \multicolumn{4}{c}{Flat} \\
 % \addlinespace
\cline{2-4} \cline{6-8}
% \addlinespace

Event time & FAR (yr$^{-1}$) & $\hat{\rho}_{T}$ & $\mathcal{M}~(\rm{M}_{\odot})$ & & FAR (yr) & $\hat{\rho}_{T}$ & $\mathcal{M}~(\rm{M}_{\odot})$ \\
% \addlinespace
\hline
\addlinespace
1238454334.99 & 114.96 & 9.12  & 0.45  &  & 93.19 & 8.79 &  0.44  \\ 
1238505374.91 & 150.64 &  8.96 & 0.43 & & -- & -- & -- \\
1238716287.79 & 177.60 & 8.95 & 0.47  & & -- & -- & -- \\
1238180069.99 & 194.74 &  9.48 & 0.39  & & -- & -- & -- \\
1238507480.87 & 213.05 & 8.98 & 0.24  & & -- & -- & -- \\
1238336288.65 & 285.36 & 9.06 & 0.51 & & 353.91 & 8.65 & 0.51 \\
1238692438.11 & 719.68 & 8.69 & 0.38 & & 170.31 & 8.74 & 0.38   \\
1238336099.82 & 557.56 & 8.73 & 0.41  & & -- & -- & -- \\
1238593683.47 & -- & -- & -- & & 303.01  & 9.39 &  0.54\\
1238204157.97 & 725.68 & 8.6 & 0.28  & & -- & -- & -- \\
\addlinespace
\hline
\end{tabular}
\label{table:candidates_chunk1}
\end{table*}

\section{Application to subsolar mass search}\label{sec:search_setup}

We perform SSM search on publicly available datasets using a two-stage hierarchical approach, as outlined in Sec.~\ref{sec:method}. The data consists of approximately five days of coincident observations from the O3 run of the two LIGO detectors—LIGO Hanford and LIGO Livingston, covering the period from April 1 to April 8, 2019.

To begin, we first conduct a coarse search over the data sampled at 512 Hz using the coarse bank described in Sec.~\ref{sec:subsolar_mass_template_bank}. The templates are generated at 35 Hz using the \texttt{TaylorF2}~\citep{spintaylor} waveform model with phase corrections up to 3.5 post-Newtonian order. The lengths of these templates range between $10^2$ and $10^3$ at 35 Hz. To prevent the templates from wrapping around during the Fast Fourier Transform operation in the matched filtering step, we ensure that the data segment length exceeds that of the longest template in the bank. Consequently, we set the data segment length to 2048 seconds for our analysis. 

We identify triggers with a matched filter SNR and re-weighted SNR~\citep{bruceallen,usman_2016} of 3.5 or greater. This threshold is chosen to increase the likelihood of detecting true signals that might be missed due to lower data sampling and the use of a coarse bank. To reduce the impact of short-duration glitches in the data, the triggers are further weighted using a chi-square and sine-Gaussian vetos. The surviving triggers then undergo a coincidence test, where they are shifted in time by 5,000 seconds, and a ranking statistic ($\Lambda$)~\citep{nitzphasetd,hlv_gareth_pycbc} is computed.

In the next step, we perform a finer search in the second stage on coincident triggers with $\Lambda \ge 7$. During this stage, matched filtering is conducted again on data sampled at 2048 Hz, using a union of nbhds around the identified coarse templates. The data sampling rate is increased by a factor of 4 to improve the matched filter SNR. Triggers from this stage are collected if their SNRs and re-weighted SNRs exceed a threshold of 4. These triggers are then re-weighted using chi-square and sine-Gaussian vetoes before undergoing a coincidence test, over the same time-shift interval as in the first stage. Finally, a list of foreground and background triggers is compiled, and the FAR for the foreground triggers is calculated following the procedure described in~\cite{hierarchical_bg_2024}.

The primary differences between the SSM search using the hierarchical method and the search adopted by the LVK collaboration~\citep{subsolar_lvk_o3a} lie in two key aspects: the parameter space covered by the template banks and the lower frequency cutoff for matched filtering. In this work, we use a coarse bank and a nbhd bank specifically designed to optimize the detection of SSM compact objects by covering a more targeted and dense parameter space. Additionally, while the flat search in~\cite{subsolar_lvk_o3a} typically starts matched filtering at 45 Hz, our approach begins at a lower frequency of 35 Hz. This choice enhances the sensitivity of our search to potential GW signals, particularly those that might be present at lower frequencies. By adjusting the matched filtering start frequency, we aim to improve the detection capabilities for signals that might be overlooked in the flat search.

\subsection{Results}\label{sec:results}

\begin{figure}[ht!]
    \centering
    \includegraphics[width=0.48\textwidth]{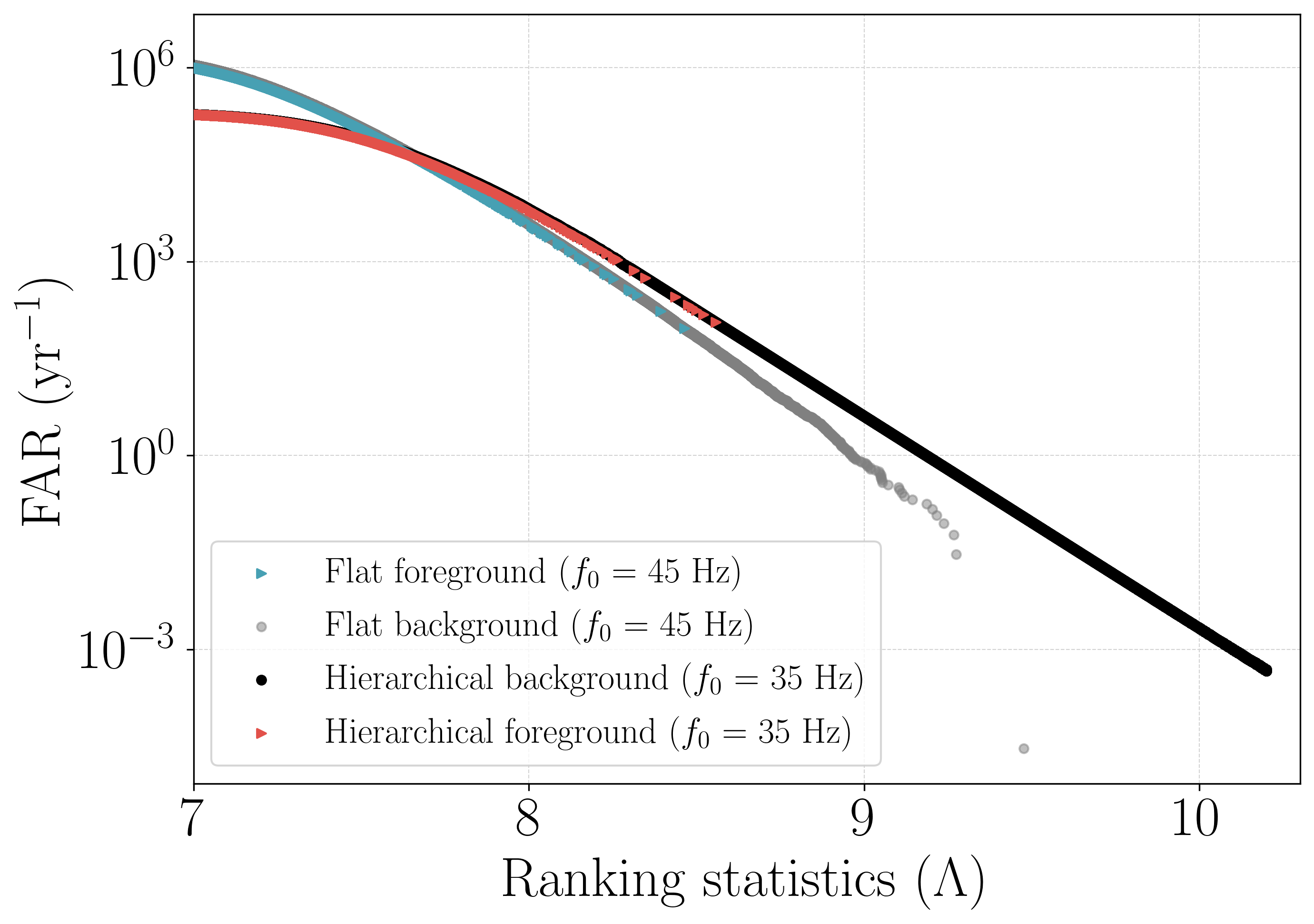}
    \caption{False alarm rate (FAR) versus ranking statistics for foreground and background computed from our reference flat and hierarchical searches. The flat search background, represented by the black curve, is computed with a time shift interval of 0.1 s using a template bank with $f_0 = 45$ Hz. The hierarchical search background, shown by the gray curve, uses the method proposed in~\cite{hierarchical_bg_2024} and a bank from a union of nbhds where templates are generated at $f_0 = 35$ Hz.}
    \label{fig:backgrounds}
\end{figure}

The hierarchical search yielded a list of GW candidates, many of which were statistically insignificant due to their FAR values exceeding 1 per year. These candidates along with the ones identified through a flat search using the same dataset are summarized in Table~\ref{table:candidates_chunk1}. While a few candidates are common to both search pipelines, none are statistically significant. As shown in Fig.~\ref{fig:backgrounds}, the foreground events overlap with the background distributions for both searches, indicating that the candidates are primarily noise coincidences.

Figure~\ref{fig:backgrounds} also highlights the disparity in backgrounds between hierarchical and flat searches, with the hierarchical search producing a larger background. This difference is due to the usage of different numbers of templates within their respective search pipelines. As shown in Table~\ref{table:table_subsolar_banks}, the coarse bank is approximately 1.6 times and the fine bank is 4.8 times denser than the flat bank owing to the template generation at 35 Hz. If the flat search had employed the fine bank, its background distribution would likely resemble that of the hierarchical search, especially in the tail. Although the number of templates used in the second stage search is low ($\sim 10-1000$ per data segment), the background generated in the second stage is expected to increase leading to more instances of noise coincidence. However, this also improves the chances of detecting GW sources that might be missed in a flat search.

\subsection{Sensitivity and Search Efficiency}\label{sec:result2}

The sensitivity of a search is determined by how many signals it can detect at a particular significance level within a given observation time (T). This can be quantified by estimating the observable volume-time (VT) product~\citep{tiwari_vt}. For a constant merger rate of the population of binaries, the average VT sensitivity product is given by 

\begin{equation}
    \langle VT \rangle = V_0 \frac{N_{\rm det}}{N_{\rm inj}} T\,,
\end{equation}

where $N_{\rm det}$ is the number of detected sources in the search, and $N_{\rm inj}$ is the total number of injected sources. $V_0$ is the volume defined as 

\begin{equation}
    V_0 = \int^{z_{\rm max}}_{0} \frac{d V_c}{dz} \frac{1}{(1+z)} dz\,,
\end{equation}

$\frac{d V_c}{dz}$ is the differential comoving volume in an expanding universe with redshift z. 

To test the sensitivity of our search method, we conducted a comparative analysis through an injection campaign on a simulated binary population. In this population, we assumed that one of the compact objects has a mass below a solar mass, while the other ranges from 1 to 10 solar masses. We created three distinct sets of injections, each defined by different spin conditions: high spin, low spin, and a mixed case where only one compact object has low spin. The distributions and ranges of the component masses and spins for these three scenarios are detailed in Table~\ref{table:table_subsolar_injection_set}.

\begin{table}[ht!]
\caption{Overview of three injection sets focusing on one of the component masses in SSM ranges. These ranges are specifically chosen to align with the parameters explored in the SSM search conducted by PyCBC~\citep{subsolar_lvk_o3a}. Each injection set has component masses (in the detector frame) and spin parameters uniformly distributed.}
\centering
\begin{tabular}{ l@{\hspace{10pt}}
c@{\hspace{10pt}} c@{\hspace{10pt}} r@{\hspace{5pt}}}
\hline
% \addlinespace
Injection set & Parameter  & Range & Waveform \\
% \addlinespace
\hline
\addlinespace
1 & $m_{1}$   & 5.0--10 M$_\odot$ & \texttt{SpinTaylorT5}\\
  & $m_{2}$  & 0.5--1.0 M$_\odot$ \\

& $\chi_{1}$, $\chi_{2}$  & 0--0.9 \\
\addlinespace
\hline
% \addlinespace
2 & $m_{1} $  & 0.2--1.0 M$_\odot$ & \texttt{SpinTaylorT5} \\
 & $m_{2} $  & 0.2--0.5 M$_\odot$ \\
& $\chi_{1}$, $\chi_{2}$  & 0-0.1 \\
\addlinespace
\hline
% \addlinespace
3 & $m_{1}$  & 0.5--5.0 M$_\odot$ & \texttt{IMRPhenomD}\\
& $m_{2}$  & 0.5--1.0 M$_\odot$ \\
& $\chi_{1}$  & 0--0.9 \\
& $\chi_{2}$  & 0--0.1 \\
\addlinespace
\hline
% \addlinespace
\end{tabular}
\label{table:table_subsolar_injection_set}
\end{table}

%/home/ksoni01/work/proj_subsolar/o3_chunk_runs/o3a/chunk1/old_runs/for_thesis_revision/new_results/calc_new_vts/vts_with_fit
\begin{figure*}[ht!]
    \centering
    \includegraphics[width=\textwidth]{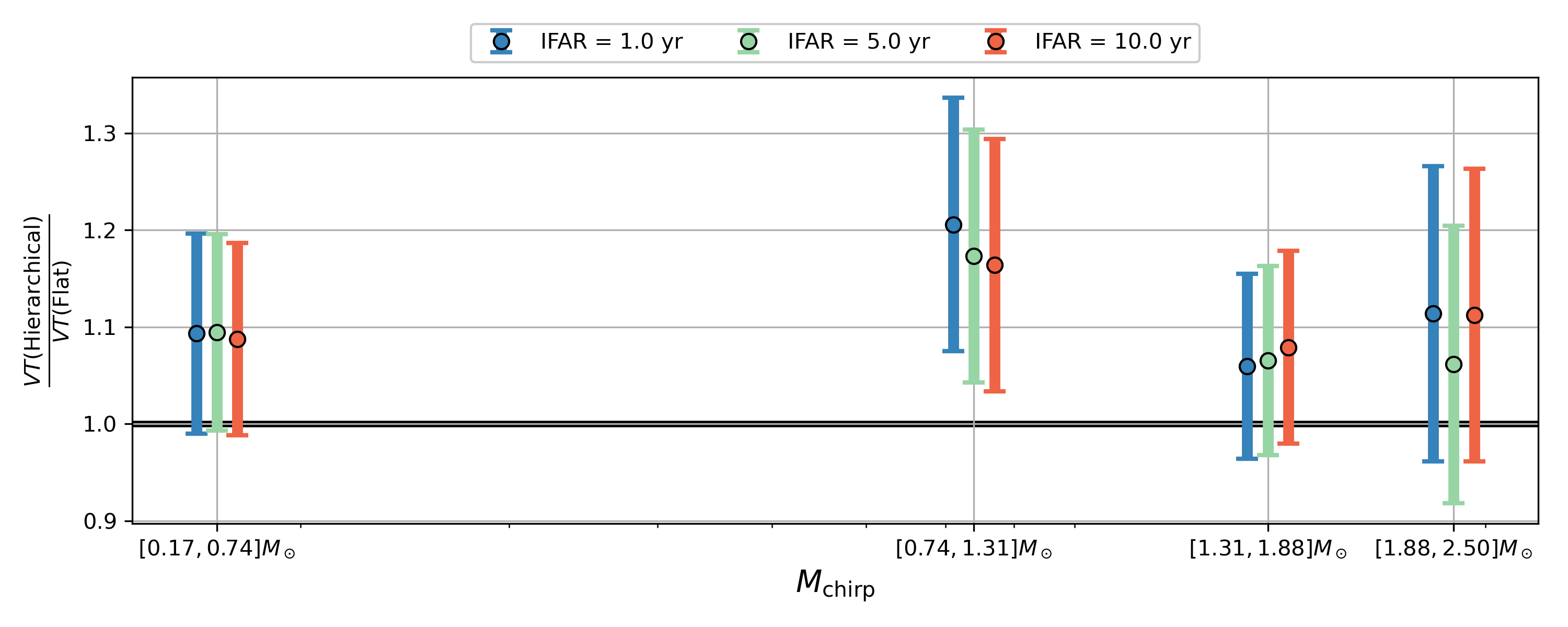}
    \caption{Plot showing the sensitive volume-time (VT) ratio for the hierarchical and flat searches, averaged across all three injection sets (see Table~\ref{table:table_subsolar_injection_set}). The VT ratio, binned over inverse false alarm rates (IFARs), demonstrates an improvement in search sensitivity across all chirp mass ($M_{\rm chirp}$) bins.}
    \label{fig:vt_chunk1_o3_subsolar}
\end{figure*}

For the given parameter space, we generated GW signals using the waveform approximants listed in Table~\ref{table:table_subsolar_injection_set}, starting at a frequency of 35 Hz. Each signal was injected into the data with a minimum interval of 100 seconds between injections, assuming an isotropic distribution for the sky locations of the sources. Following this procedure, approximately 6,500 injections were made across the three sets, and a search was performed using both the traditional flat method and our hierarchical approach.

Figure~\ref{fig:vt_chunk1_o3_subsolar} shows the VT ratio computed for the two search methods. The hierarchical search outperforms the flat search, showing an improvement in the VT ratio by approximately 1.1 to 1.2. This improvement is primarily due to the SNR enhancement when the search is performed at a lower frequency of 35 Hz instead of 45 Hz. When the search is performed at 35 Hz, we expect the gain in the SNR to be approximately 6\% and an astrophysical volumetric gain of $\sim20\%$. However, the injection study results indicate that the detection volume improves by only $10-20\%$. This discrepancy likely arises from an increase in the noise background when the lower bound of the matched filtering frequency is reduced.

%However, through the injection study, we see that the detection volume could improve by only $10-15\%$. This could happen because the noise background also becomes higher as we reduce the lower limit of the matched filtering frequency. 

%http://localhost:8888/notebooks/Documents/mount_ogrid/work/proj_subsolar/o3_chunk_runs/o3a/chunk1/vt_calc/speed_up_cal.ipynb
\begin{table}[ht!]
\caption{Comparison of CPU core hours required for the flat search and the coarse and fine stages of the hierarchical search. The values outside the parentheses represent the CPU core hours for the Hanford detector, while the values inside the parentheses correspond to the Livingston detector.} 
\centering
\begin{tabular}{ c@{\hspace{30pt}} c@{\hspace{5pt}}}
\hline 
Search & CPU core hour \\ 
\hline
\addlinespace
Flat & 30603.19 (30163.08) \\
Coarse & 11205.94 (11357.42) \\
Fine & 1116.18 (1115.68) \\
\addlinespace
\hline
\end{tabular}
\label{table:cost}
\end{table}

We compared the matched filtering cost by comparing the number of CPU core hours required by each detector in use in the two searches. As shown in Table~\ref{table:cost} the number of CPU core hours required by flat search is more than coarse and fine searches owing to the different number of templates used in each search. The numbers show that the overall cost of a hierarchical search is approximately 2.5 times less than a flat search. This is a huge advantage of hierarchical search given that the search sensitivity is also more than that of the flat search.

\section{\label{sec:conclusion}Conclusion and discussion}

The search for long-duration GW signals from compact binary mergers, such as SSM binaries, low-mass BNSs, precessing binaries, and binaries with moderate mass ratios ($m_1/m_2\,, m_1 \ge m_2$) with eccentricity in their orbits, presents significant challenges in the current LIGO-Virgo frequency band. These challenges primarily arise from the requirement of large, densely populated template banks necessary for matched filtering. To mitigate the computational burden, suboptimal choices are often made, which inevitably limit the sensitivity of such searches. With the advent of 3G detectors, these challenges are expected to become more pronounced, as GW signals will be observed over longer durations, ranging from tens to hundreds of minutes. This extended observation window will substantially increase the computational demands due to the rapid expansion of the parameter space. Consequently, the development of efficient hierarchical search strategies is critical, not only to enhance current detection capabilities but also to ensure readiness for the vastly more computationally expensive searches required by next-generation detectors.

In this paper, we demonstrated that a hierarchical search strategy can be effectively implemented for the search of SSM binaries without restricting the parameter space. In Sec.~\ref{sec:method}, we presented a preliminary calculation indicating that the SNR improves by approximately 6\% when the matched filtering is conducted starting at a frequency of 35 Hz. This means that a $\sim 20\%$ increase in the sensitive volume could be expected in the search. Building on the results from Sec.~\ref{sec:method}, we constructed the necessary template banks (see Sec.~\ref{sec:subsolar_mass_template_bank}) and conducted a hierarchical search on a small dataset from the O3 run (see Sec.~\ref{sec:search_setup}). As described in Sec.~\ref{sec:results}, our search did not yield any significant candidates, which is consistent with previous searches~\citep{lvk_subsolar_mass_search_2022,subsolar_lvk19,nitz_wang_subsolar,Nitz_pbh_2021,subsolar_lvk_o3a,subsolar_lvk21}. Through injection studies, we demonstrated that the hierarchical search gives a sensitive VT ratio improvement of about $10-20\%$ compared to the flat search method employed by the LVK Collaboration. This volumetric improvement is important as this can increase the possibility of finding sources in the upcoming LIGO and Virgo observation runs. Moreover, this improvement can provide better constraints on the fraction of dark matter, potentially ruling out several models proposing SSM black holes as dark matter candidates.

Our findings highlight that near-optimal sensitivity can be achieved using a hierarchical search, even when compared to a direct flat search. By optimizing different aspects of the search process at two stages, such as adjusting the frequency of operation, data sampling rates, and the density of the template banks, we achieved computational savings of up to a factor of 2.5 while simultaneously enhancing the sensitivity of the SSM search. This represents a significant advancement in search optimization as we prepare for 3G detectors.

% Looking ahead, 3G detectors will introduce several key challenges, as discussed in Sec.~\ref{sec:intro}, for CBC searches due to their increased low-frequency sensitivity. These detectors are expected to observe a significantly higher number of GW signals from a variety of CBC sources, including eccentric and precessing binaries, over extended durations. These sources would increase the dimensionality and ranges of the search parameter space thereby making the current search pipelines ineffective for their search. To manage this expanding parameter space effectively and reducing associated search computational cost, a hierarchical search approach will be crucial.

% potentially increasing the parameter space, but they will also observe signals that remain in the sensitive frequency band for extended durations. The expansion of the search parameter space, particularly due to eccentric or precessing binaries, can be effectively addressed through a hierarchical search approach. 

Looking ahead, 3G detectors are expected to introduce several significant challenges for CBC searches due to their enhanced low-frequency sensitivity, as discussed in Sec.~\ref{sec:intro}. With the ability to detect GW signals from a broader range of CBC sources—including eccentric and precessing binaries over extended durations—the search space will expand dramatically. This increase in the search space will, in turn, raise computational demands to potentially unmanageable levels. Therefore, a hierarchical approach may be proposed to address these challenges effectively.

In the 3G era, the hierarchical search could be structured into stages, with the first stage focused on efficiently identifying potential candidates by maximizing the likelihood of detecting GW signals based on all intrinsic parameters, while the second stage aims to improve the SNR and address any losses from the first stage. Since the primary goal of the first stage is to locate regions where signals are likely to be present, general optimizations related to reducing the template bank size, adjusting data sampling, and defining the operating frequency range for matched filtering can be applied, as shown in this work. The density of the template bank could be reduced by coarsening the bank and adjusting the frequency at which templates are generated. However, this step should concentrate on regions of the parameter space where the SNR due to features in the binary’s orbit is expected to be higher. For instance, in the case of eccentric binaries, the effects of eccentricity are most prominent in the lower frequency region, while at higher frequencies, the binary’s orbit is expected to circularize. Therefore, the template bank, matched filtering frequency band, and data sampling rate could be adjusted to focus only on the higher frequencies to reduce the search cost. To enhance the detection probability, a nbhd search could be performed with templates having eccentricities at lower frequencies in the second stage. A similar strategy could be applied to binaries with moderate precession. For these binaries, the first stage can be adjusted to search for signals with variable starting frequencies for matched filtering, excluding less significant merger-ringdown phases. However, these regions could be relaxed in the second stage. Since we expect the Earth’s rotation to impact the antenna response functions, potentially reducing search sensitivity, approximate response functions, which would include the effect of the source's orientation with respect to the detector, could be introduced only in the second stage of the hierarchical search, thereby improving overall sensitivity. The second stage could also be optimized to recover any SNR loss from the first stage resulting from the various optimizations applied earlier. 

% For example, the template bank density could be reduced by coarsening the bank and adjusting the frequency at which templates are generated, ensuring that the search captures most of the signal power in the low-frequency band, as demonstrated in this study. Additional optimizations, such as using variable starting frequencies for matched filtering to exclude less significant merger-ringdown phases in long-duration signals and adjusting data sampling rates, can further reduce computational complexity. Furthermore, the effects of moderate precession and eccentricity in the binary's orbit could be temporarily relaxed in the first stage to lower computational costs, with these effects fully incorporated in the second stage. In the case of long-duration signals, the Earth’s rotation may impact the antenna response functions, potentially reducing search sensitivity. To address this, approximate response functions could be introduced in the second stage of the hierarchical search, improving overall sensitivity. Additionally, the second stage could be tuned to recover any SNR loss from the first stage, resulting from the various optimizations applied earlier. 

\begin{acknowledgments}
    KS and AHN acknowledge support from the National Science Foundation grant (PHY-2309240). KS acknowledges the support for computational resources provided by the IUCAA LDG cluster Sarathi and Syracuse University through the OrangeGrid High Throughput Computing (HTC) cluster. KS expresses sincere gratitude to Sanjit Mitra for discussions in the very early stages of this work. 
\end{acknowledgments}


% \vspace{5mm}
% \facilities{HST(STIS), Swift(XRT and UVOT), AAVSO, CTIO:1.3m,


% \software{astropy \citep{2013A&A...558A..33A,2018AJ....156..123A},  
%           Cloudy \citep{2013RMxAA..49..137F}, 
%           Source Extractor \citep{1996A&AS..117..393B}
%           }



% \appendix

% \section{Appendix information}

\bibliography{references}{}
\bibliographystyle{aasjournal}



\end{document}